\documentclass[aps,prb,twocolumn,floatfix,letterpaper]{revtex4}
\usepackage{amsmath}
\usepackage{epsfig}
\usepackage{graphicx,color,floatflt,mathrsfs,float}
\usepackage[ps2pdf,a4paper=false,letterpaper=true,colorlinks=false,urlcolor=blue,bookmarks=true]{hyperref}
\usepackage[tight]{subfigure}
\usepackage{mathptmx}
\begin{document}

\title{Imaging and manipulating electrons in a 1D quantum dot with Coulomb 
blockade microscopy}
\author{Jiang Qian$^{1,2}$, Bertrand I. Halperin$^1$ and Eric J.  
Heller$^1$}

\affiliation{$^1$ Physics Department, Harvard University, Cambridge, MA 
  02138, USA\\ $^2$ Arnold Sommerfeld Center for Theoretical Physics and 
Center for NanoScience,
Ludwig-Maximilians-Universit\"at M\"unchen, Germany}
\begin{abstract}
    Motivated by recent experiments by the Westervelt group, which used 
a mobile tip to probe the electronic state of a segmented nanowire, we 
calculate shifts in Coulomb blockade peak positions, as a function of 
tip location, which we term ``Coulomb blockade microscopy''.  We show 
that if the tip can be brought sufficiently close to the nanowire, one 
can distinguish a high density electronic liquid state from a Wigner 
crystal state by microscopy with a weak tip potential.  In the opposite 
limit of a strongly negative tip potential, the potential depletes the 
electronic density under it and divides the quantum wire into two 
partitions.  There the tip can push individual electrons from one 
partition to the other, and the Coulomb blockade micrograph can clearly 
track such transitions.  We show that this phenomenon can be used to 
qualitatively estimate the relative importance of the electron 
interaction compared to one particle potential and kinetic energies.  
Finally, we propose that a weak tip Coulomb blockade micrograph 
focusing on the transition between electron number $N=0$ and $N=1$ 
states may be used to experimentally map the one-particle potential 
landscape produced by impurities and inhomogeneities.

\end{abstract}
\date{\today}
\maketitle

\section{Introduction}\label{intro}

Studies of nanoscale electronic structures hold important promise both 
as laboratories for few-body, interacting quantum mechanical systems and 
as technological testbeds for future classical or quantum computing 
technologies. Novel probe technologies ~\cite{amir_science, 
topinka_science} are very important for studying electronic properties 
in nanoscale systems because they are often beyond the resolution of 
conventional imaging techniques like optical microscopy, and traditional 
transport measurements can only measure spatially averaged physical 
properties such as the conductance or the current.  One scanning probe 
microscopy (SPM)~\cite{topinka_science} technique utilizes a charged 
metallic tip to perturb the local electronic density in a 
nanoelectronic structure while monitoring the resulting change in 
transport properties.  Using this technique one can obtain spatially 
resolved measurement of the electronic properties, including the local 
electron density and, in principle, the wavefunction itself in the case 
of a one-electron system~\cite{Fallahi}(see discussions below).  This 
imaging technique has been fruitfully applied to study the flow of 
ballistic electrons across a range of two dimensional heterostructures.

Recently, a series of experiments~\cite{ania_prepreint} applied SPM 
techniques to study quantum wires.  In these experiments a segment of 
an InAs nanowire lying on top of a two-dimensional SiO$_x$ layer was 
isolated from the rest of the wire by two short InP segments, forming a 
one-dimensional quantum dot with lithographically defined boundaries.  
A negatively charged probe scanned controllably the two dimensional 
area around the wire and the  conductance across the 1D quantum dot was 
measured as a function of the probe location.  Both the voltage of the 
probe and its height above the surface can also be independently 
varied.  Motivated by these new experimental possibilities, we turn to 
exact diagonalization techniques to study the conductance response of a 
few-electron quantum dot as a function of a spatially varied probe 
potential, in order to illustrate the kind of information that can be 
extracted in the case of a system of several electrons.

\section{Model}\label{sec:model}
\begin{figure}
\centering
\includegraphics[width=0.40\textwidth]{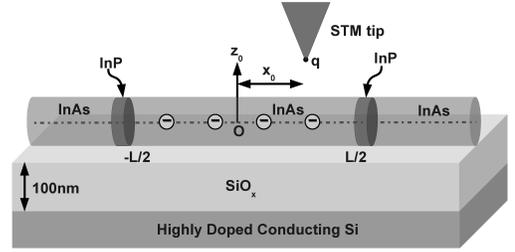}
\caption{\label{fig:geometry}Schematic geometry of Coulomb blockade 
microscopy of a quantum wire containing four electrons. When calculating 
the electron-electron interaction and the electron-tip interaction, we 
assume that the InP barriers have zero thickness, the InAs wire is 
infinitely long, and the substrate layers extend to infinity in x and y 
directions.}
\end{figure}
\begin{figure}
\centering
    \includegraphics[width=0.48\textwidth]{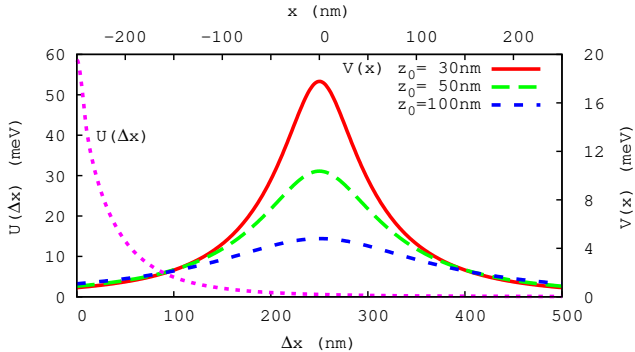}
    \caption{\label{fig:tip}(Color Online) (Dotted)Interaction potential 
$U(\Delta z)$.(Solid, long and short dashed) Tip $V(x)$
potentials with a tip charge $q=e$ and locations $\vec{r_0}=(0,0,z_0)$ 
where $z_0=30nm,50nm,100nm$.}
\end{figure}
We consider a uniform InAs (dielectric constant $\epsilon=15.4$, Bohr 
radius $a_B\approx34nm$) nanowire of radius $R=10$nm, which lies in 
vacuum atop a SiO${_x}$ ($\epsilon=3.9$) layer $100$nm thick, 
separating it from conducting doped bulk silicon~(see 
Fig.~\ref{fig:geometry}).  Electrons are modelled as point charges 
traveling
along the center axis of the wire, confined to interval 
$-\frac{L}{2}<x<\frac{L}{2}$ by hard walls, representing the InP layers.  
We consider length $L$ from $110$nm to $500$nm. The electron-electron 
interaction $U(x_1-x_2)$ was calculated using the commercial 
finite-element program Comsol\texttrademark to solve the classical 
Poisson equation for a point charge on the axis of an infinite wire 
above in a substrate with the geometry described in 
Fig.~\ref{fig:geometry}. At short distance $\Delta x$, the potential was 
softened to account for the finite thickness of the electron 
wavefunction, by replacing $\Delta x^{-1}$ with $[(\Delta 
x)^2+R^2]^{-1/2}$. Following the approximation used by 
Topinka~\cite{topinka_thesis}, we model the negatively charged probe as 
a fixed point charge of strength $q$ at a location $\vec{r_0}$ relative 
to the center point of the wire. This gives rise to a one-body potential
$V(x;\vec{r_{0}},q)$ for an electron on the wire axis at point $x$, which we 
again obtain by solving the Poisson equation~(results are shown in 
Fig.~\ref{fig:tip}).
 
In this paper, we diagonalize the exact 1D many-body Hamiltonian with 
the Lanczos method~\cite{lanczos} for up to electron number $N=4$:
\begin{equation}\label{equ:Hamiltonian}
    -\frac{\hbar^2}{2m^*}\nabla^2\Psi+\sum_{i=1}^{N}V(x_{i};\vec{r_0},q)\Psi
    +\sum_{i=1}^{N}\sum_{j=1}^{i-1}U(x_i,x_j)\Psi
    =E~\Psi,
\end{equation}
where $\Psi$ is the full many-body wavefunction, depending on the 
position $x_i$ and spin $\sigma_{i}$ of the electrons.  To connect to 
the experimentally observable variables, we consider the Coulomb 
blockade peak $\emph{positions}$ of the transition from $(N-1)$ to $N$ electron 
ground states. The conductance through the quantum wire is maximum when the 
chemical potential difference between the lead and the wire, controlled by the
voltage $V_g$ on a back gate, is equal to the ground state energy difference 
between the two states in question.  We may write this condition as $\Delta E 
\equiv E_{N} - E_{N-1} = \alpha V_g + \beta $, where $\beta$ is a constant and 
$\alpha$ is the proportionality constant between changes in the back gate 
voltage and the chemical potential in the quantum dot. We probe the electronic 
states in the quantum wire through the dependence of $\Delta E$ on the tip 
position $\vec{r_0}$ and potential strength $q$. An interesting set of 
spatially resolved information about the electrons in the wire can be extracted 
from this function, and we call this method ``Coulomb blockade microscopy''. It 
is a special application of the ``scanning probe microscopy'' developed by the 
Westervelt group~\cite{topinka_science}.  In calculations in this paper we 
focus on the transition from $N=3$ to $N=4$ electrons, but most of our 
conclusions are easily generalizable to other ground state transitions.  

Finally, we note that for four non-interacting electrons with spin in a 
wire of radius $R=10nm$, when the dot length $L>L_m=18.2~nm$, the 
lowest four single particle energy levels are all longitudinal modes.  
The shortest wire length we consider in this paper $L=110nm\gg L_m$, we 
therefore expect the wires under consideration can be well approximated 
as strictly 1D under the assumption of weak interaction effects.  
Indeed, the gap between the transverse ground state and first excited 
state for our wire is $\Delta E\approx 148meV$.  The most confined 
geometry we discuss is shown in Fig.~\ref{fig:length_strong_tip}: a 
$L=110nm$ wire with four electrons under an extremely strong tips, 
squeezing them into the both ends of the wire. Even in that case, the 
\emph{total} energy per electron is less than $140meV$, not enough to 
cause an excited transverse mode, with full interaction effects taken 
into account. Thus we expect the quantum wire to be well approximated 
by a 1D model for all the parameters we explored in this paper.

The existence of higher transverse modes will lead to a renormalization
of the effective electron-electron interaction (generally a softening of
the potential at short distances) even if the energies of the modes are
well above the Fermi energy. This is because two electrons that come close
together in the lowest transverse mode will mix, virtually, with states
where the electrons are in excited modes. As our calculations use, already,
a crude phenomenological cut-off at short distances, we do not include
explicitly effects of this renormalization. Technically, the existence of
higher transverse modes can also lead to three-body and four-body effective
interactions, arising from three-body and four-body collisions, but we do
not expect such higher body terms to be important in the wires under
consideration.

We want to emphasize that although we choose a specific wire geometry  
and a specific form of interaction that model some characteristics of 
the setup of the Westervelt group's on-going experiments, most features 
of the Coulomb blockade micrographs we discuss below are applicable to 
any 1D quantum dot system under a mobile potential. Indeed, our 
discussions of the qualitative features of Coulomb blockade micrographs 
under both the weak-tip and the strong-tip limits rely only on the 
general properties of 1D electronic systems, independent of the 
specific geometry and interaction we adopt. Our numerical results 
mostly serve for illustrative purposes.

\section{Electronic Density}\label{density}
\begin{figure}
\centering
    \includegraphics[width=0.40\textwidth]{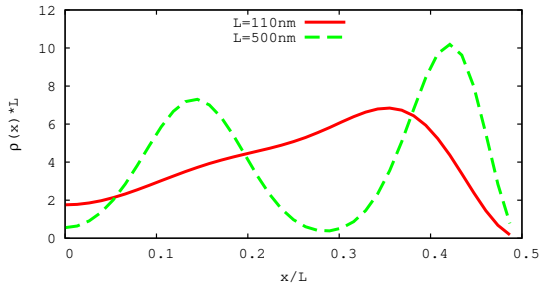}
\caption{\label{fig:length_density}(Color Online) Electronic densities for a 
$L=110nm$ wire and for a $L=500nm$ wire in absence of tip potential.  
Only the right half is shown, as the plot is symmetric about $x=0$. We 
rescale the $x$ and $\rho(x)$ with wire length $L$ to ease the 
comparison.}
\end{figure}
In the absence of a probe potential $V$, both the $N=3$ and $N=4$ wires 
the electronic density profile $\rho(x)$ undergoes a crossover as a 
function of $L$ from a liquid state characterized by a $2k_F$ Friedel 
oscillations to a quasi-Wigner crystal state characterized by a $4k_F$ 
density oscillation. Such a crossover from a liquid state to 
quasi-Wigner states with a decrease in density a very generic 
phenomenon for 1D interacting fermion 
system~\cite{greg_prb,matveev_prb}. Indeed, for any interacting 
decaying no faster than $x^{-2}$ at long distance a quasi-Wigner 
crystal state is known to emerge at low density~\cite{matveev_prb}.  
This requirement for interaction will hold for a system with long 
screening length as compared to mean inter-particle distance, as is the 
case for our geometry when the screening doped silicon layer is 
relatively far ($100\mu m$) away. For our specific geometry and 
interaction, the crossover happens at around density $\rho^*\approx 
35\mu m^{-1}$.  A Wigner crystallized density variation is shown in the 
dashed curve of Fig.~\ref{fig:length_density} for a quantum dot of 
$L=500nm$, whereas for $L=110$ the four electron density exhibits 
Friedel oscillations.

\section{Weak Tip Limit}\label{weak}
\begin{figure}
\centering
    \includegraphics[width=0.40\textwidth]{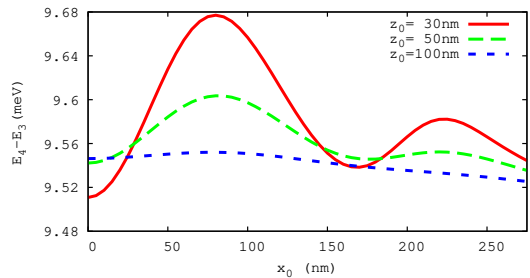}
\caption{\label{fig:height_gap}(Color Online) Coulomb blockade micrographs for 
a 1D dot with $L=500$nm and tip charge $q=0.02e$ for three tip potential 
shown in Fig.\ref{fig:tip}. Again, the right half is shown.}
\end{figure}
\begin{figure}
    \includegraphics[width=0.40\textwidth]{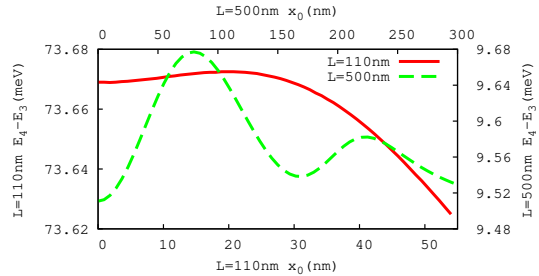}
\caption{\label{fig:length_gap}(Color Online) $q=0.02e$ weak tip limit 
for the $N=3$ to $N=4$ Coulomb blockade transition for densities shown 
in Fig.~\ref{fig:length_density}. The tip distance to the wire is 
$z_0=30nm$. Only the right half is shown.}
\end{figure}
\begin{figure}
    \includegraphics[width=0.40\textwidth]{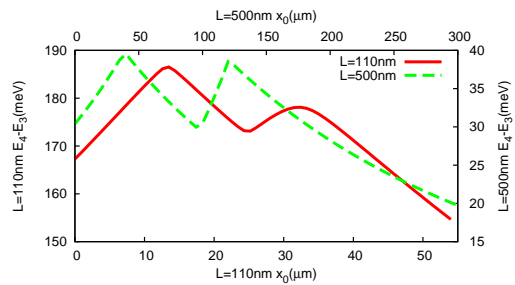}
\caption{\label{fig:length_strong_tip}(Color Online) $q=8e$ strong tip 
limit for the $N=3$ to $N=4$ Coulomb blockade transition for densities 
shown in Fig.~\ref{fig:length_density}  The tip to wire distance is 
$30nm$.  Only the right half is scanned.  }
\end{figure}

Now we introduce a weak tip potential, corresponding to a negatively 
charged tip of strength $q=0.02e$, scanning above the center axis of the 
quantum wire along its direction $(1,0,0)$, with the tip location vector 
$\vec{r_0}=(x_0,0,z_0)$.  For a 1D quantum dot of length $L=500nm$, 
which as shown in Fig.~\ref{fig:length_density} has $4k_F$ 
Wigner-crystal density variation, let us consider the three tip heights 
above the quantum wire, $z_0=30nm, 50nm, 100nm$, corresponding to the 
three tip potential shown in Fig.~\ref{fig:tip}.  The resulting Coulomb 
blockade peak position $\Delta E$ as a function of the tip coordinate 
$x_0$ along the wire, i.e.  the Coulomb blockade micrograph, is shown in 
Fig.~\ref{fig:height_gap}.  Clearly, in Fig.~\ref{fig:tip} the closer 
the tip approaches the wire, the more localized is the tip potential and 
a sharper tip potential make it easier to resolve the density 
variations, this is reflected in Coulomb blockade micrograph scans in 
Fig.~\ref{fig:height_gap}.  At $z_0=30nm, 50nm$ from the tip to the 
center of the wire, the $4k_F$ density oscillation of the quasi-Wigner 
crystal state on the right can be detected in the Coulomb blockade 
micrograph, whereas when $z_0=100nm$ away, the tip potential becomes 
much too broad to resolve the fine features of the density oscillations.  
We note that although the resolution of the tip is largely determined by 
the distance $z_0$, the contrast of a Coulomb blockade micrograph, i.e.  
the magnitude of the $4k_F$ variations in the micrographs, can be 
improved by modestly increasing the tip potential.

By contrast, in Fig.~\ref{fig:length_gap} the $L=110$nm micrograph at 
$z_0=30nm$ does not show features of Wigner crystal oscillations.  
However, this micrograph does not by itself give a clearcut indication 
of the absence of Wigner crystal order for $L=110$nm. With the current 
interaction and tip parameters, one cannot observe the crossover from 
the Wigner crystal to the  Friedel oscillations because it happens at a 
inter-particle spacing $\Delta x\approx30nm$, below the resolution of 
the micrograph even at $z=30$nm. To be more specific, we may define the 
onset of Wigner-crystal order for our four-electron system as the 
length $L$ at which there first appears a local minimum of the mean 
density $\rho(x)$ in the vicinity of $x/L = 0.25$.  According to our 
calculations, this should occur at L=135 nm. However, with the tip at 
height 30nm, in the weak charge limit, the resolution of the micrograph 
is of the order of 60nm,  so we would not see a secondary  minimum in 
the micrograph signal until $L \geq 250 nm.$

To gain a more intuitive understanding of a weak tip Coulomb blockade 
micrograph, we observe that a weak tip only slightly disturbs the 
electron density as it scans across the wire, thus, a simple first order 
perturbation theory should be a good approximation to compute the ground 
state energy in the presence of the tip potential:
\begin{equation}
\label{equ:perturbation}
    E(\vec{r},q)-E_0(\vec{r},q)
=\int dx~V(\vec{r_0};q,x) \rho(x),
\end{equation}
where $\rho(x)$ is the non-interacting  ground state density and 
$E_0(\vec{r_0},q)$ is its energy. We have checked that for tip charges 
up to $q=0.1e$ the simple first order perturbation theory gives a decent 
fit to both the ground state energy and the Coulomb blockade micrograph.
Since both the width and the center location of the tip potential 
$V(\vec{r_0};q,x)$ can be adjusted experimentally,  the Coulomb blockade 
microscopy with a weak tip potential provides a flexible way to map the 
electronic densities in a quantum dot. Indeed, as an example of such 
flexibility, we find that one can improve the ``contrast'' of a Coulomb 
blockade micrograph; i.e.,  the prominence of the spatial variations in 
micrographs like Fig.~\ref{fig:height_gap} as compared with the total 
energy shift $E_4-E_3$, can be improved by slightly increasing the tip 
potential while still staying within the weak tip perturbative 
approximation.


\section{Strong Tip Limit}\label{strong}
\begin{figure}
    \includegraphics[width=0.48\textwidth]{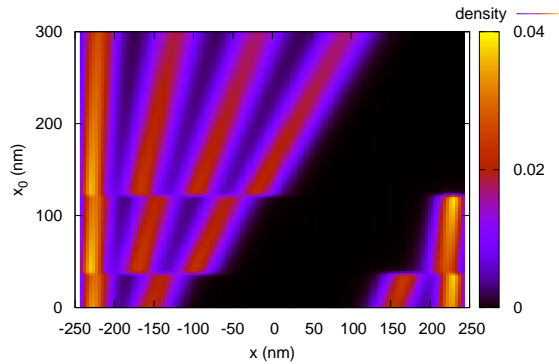}
\caption{\label{fig:strong_tip_density}Electronic densities in 
$L=500nm$, $N=4$ quantum wire as a tip $q=8e$ scan through $0\leq 
x_0\leq 250nm$. The density throughout the entire wire $-250nm\le x\le 
250nm$ is shown.}
\end{figure}
In the opposite limit of strong tip, the Coulomb blockade tip scans 
present a very different physical picture.  In 
Fig.~\ref{fig:length_strong_tip} we observe that irrespective of 
whether the electronic state is liquid or Wigner-crystal like as shown 
in Fig.~\ref{fig:length_density}, the Coulomb blockade micrographs show 
similar behavior: in the case of $N=3$ to $N=4$ transition, both the 
$L=110$nm and $L=500$nm wire show two relatively sharp peaks for a 
large tip charge $q=8$.  This is in contrast with the case of a weak 
tip Fig.~\ref{fig:length_gap}, where the Coulomb blockade micrographs 
show smooth spatial dependence as well as sensitivity to the electronic 
states in the absence of the tip potential.


To understand the physics of this strong tip limit we note that the two 
sharp cusps in Fig.~\ref{fig:length_strong_tip}  represents 
discontinuous slope changes in the $N=4$ electron ground state energy as 
a function of tip position $x_0$. Similarly the deep valley in the 
figure corresponds to a cusp in $N=3$ ground state energy. The origin of 
these three discontinuities in slopes can be seen in 
Fig.~\ref{fig:strong_tip_density}.  In this limit, the negatively 
charged tip potential is so strong that it depletes the electronic 
density under it.  Thus the tip creates an effective partition of the 
electrons in the wire into left and right sub-quantum dot. As shown in 
Fig.~\ref{fig:strong_tip_density}, as the tip move from the center to 
right of the wire with four electrons, the partitions of the electrons 
undergoes two abrupt transitions $(2,2)\to(3,1)\to(4,0)$.  These two 
transitions correspond  to the two upward cusps shown in the $N=4$ 
curve in Fig.~\ref{fig:length_strong_tip}. Similarly, the discontinuous 
slope change shown on the $N=3$ curve of the 
Fig.~\ref{fig:length_strong_tip} corresponds to the transition between 
the $(2,1)\to(3,0)$ partition of the ground state.  Thus the three 
discontinuities seen in the Coulomb blockade micrographs in 
Fig.~\ref{fig:length_strong_tip} correspond to, alternately, the 
transitions between the integer partitioning of total electron numbers 
in the $N=3$ and $N=4$ system. The upward slope of the curve near 
$x_0=0$ reflects an additional downward cusp at the origin, due to the 
transition (1,2)$\to$(2,1) in the $N=3$ wire.

\begin{figure}
\centering
\includegraphics[width=0.5\textwidth]{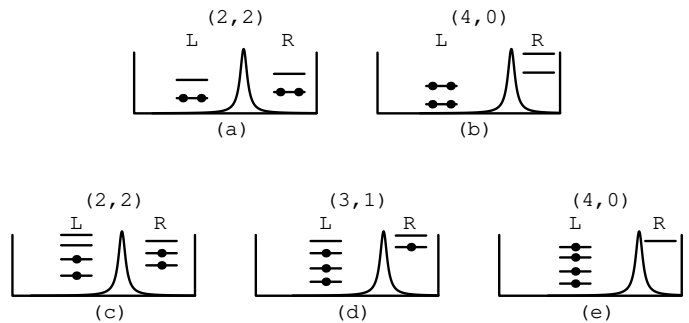}
\caption{\label{fig:strong_schematic} Schematic illustrations of partition of 
the four-electron state by the potential barrier under a strong tip
potential. In (a) and (b), single-electron energy levels are shown for
non-interacting electrons, for two positions of the tip. Because of spin 
degeneracy, we find either two electrons in each well, or all four in
the same well, when the energy of the single-electron ground state on
the right becomes higher than the first excited level on the left. In
(c)-(e), we ``incorporate'' interaction energy into ``single electron
levels'' schematically by plotting the energy needed to add an additional 
electron. In drawing energy levels this way, it is clear that the spin
degeneracy in the non-interacting case is lifted by electron interaction, and 
an additional (3,1) partition will appear for the ``energy level''
arrangement in (d). Filled circles show electrons in occupied 
levels.}
\end{figure}

To better understand the transitions between different partitions, let 
us consider the transitions in a model of $N=4$ electrons, with spin, 
which have no Coulomb repulsion between them but interact with a repulsive tip 
potential. The scenario is illustrated in Fig.~\ref{fig:strong_schematic}, 
panels $(a)-(b)$. When tip is near the center of the wire, the electrons are 
partitioned $(2,2)$ and both electrons in each side reside in the 
single-particle ground state.  As the tip moves rightward, the energy levels 
rise in the right partition and fall in the left.  When the first excited level 
on the left partition crosses the ground state on the right, $\emph{both}$ 
electrons will move to the left partition.  Therefore, contrary to the 
interacting case, there is no energetically favorable state of $(3,1)$ 
partitioning in the non-interacting system.  In the $N=3$ case, the $(2,1)$ 
partition is not affected by this, and for non-interacting system the 
transition $(2,1)\to(3,0)$ will coincide with the transition in $N=4$, so the 
Coulomb blockade micrograph will show only a single peak. This analysis can 
also be generalized to a wire containing multiple non-interacting electrons, 
such that  all the $(odd,odd)$ partitionings of electron number will be 
missing.

As schematically illustrated in Fig.~\ref{fig:strong_schematic}, panels 
$(c)-(e)$, when we take electron interaction into consideration, the one and 
two electron state would no longer be degenerate in either partition, so 
contrary to the non-interacting scenarios in panel $(a)-(b)$, here a $(3,1)$ 
partition can survive as an intermediate stage between the $(2,2)$ and $(4,0)$ 
partitions.  With the non-interacting case in mind, we postulate that the 
distance between two peaks in $N=3$ to $N=4$ micrographs, corresponding to the 
tip positions where $(3,1)$ partitioning in the $N=4$ wire is stable, can serve 
as an indicator of the relative importance of the interaction energy versus the 
sum of kinetic and single particle potential energies. The less important 
interaction is compared to single particle energies, the less splitting would 
the one and two particle energies be, and the smaller is the region of stable 
$(3,1)$ partition. This can be seen in Fig.~\ref{fig:length_strong_tip}. The 
potential energy should have a larger share in the total energy in the longer 
wire with lower electronic density, and indeed we observe that the longer wire  
has a wider distance between the two peaks marking $(2,2)\to(3,1)$ and 
$(3,1)\to(4,0)$ transitions.

\section{Intermediate Tip Charge}\label{intermediate}
\begin{figure}
\includegraphics[width=0.40\textwidth]{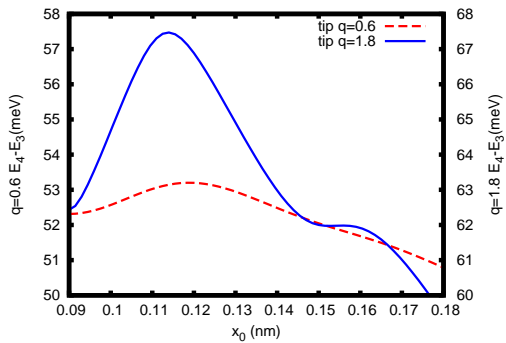}
\caption{\label{fig:micrograph_tip} (Color Online). $q=0.6e$ and 
$q=1.8e$ intermediate tip potentials for $N=3$ to $N=4$ Coulomb 
blockade transition in a wire of length $L=180nm$. The tip distance to 
the wire is $z_0=30nm$. Only the right half is shown.}
\end{figure}

\begin{figure}
\includegraphics[width=0.40\textwidth]{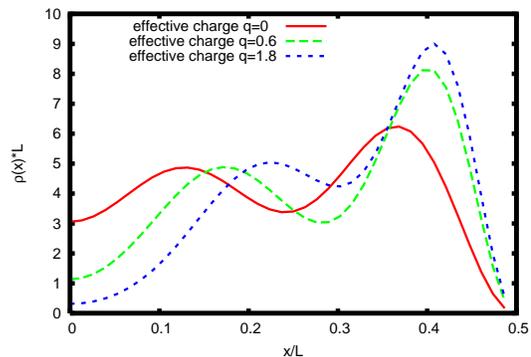}
\caption{\label{fig:density_under_tip}(Color Online) Electronic density 
for a $L=180nm$ wire under three tip potentials, with effective charges 
$q=0e$, $q=0.6e$ and $q=1.8e$. Only the right half is shown, as the 
plot is symmetric about $x=0$. We rescale the $x$ and $\rho(x)$ with 
the wire length $L$.}
\end{figure}

We have carried out calculations with various tip charges intermediate
between the two limits discussed above.  In general, the larger the charge
on the tip, the more readily one sees the secondary minima in the
micrograph signal, which are seen in Fig.~\ref{fig:length_strong_tip} for a 
charge $q=8e$, even at $L=110 nm$.  As one illustration, for a wire of length 
180 nm, with a tip height of 30 nm, we find that the micrograph signal shows a 
secondary minimum at $x/L \approx 0.15$ for $q=1.8e$, but shows no 
secondary minimum when $q=0.6e$, as seen in 
Fig.~\ref{fig:micrograph_tip}  However, if we calculate the electron 
density in the wire when the tip is over the center of the wire ($x_0 = 
0$),  we find that a
tip charge of $q=0.6e$ is enough to substantially modify the density
relative to the  density in the absence of the tip. As seen in 
Fig.~\ref{fig:density_under_tip}, the electron density below the tip, 
at $x=0$ is reduced by a factor of three relative to the density with 
no tip charge.  Nevertheless, the oscillations seen in $\rho(x)$  
remain qualitatively similar to structure seen in the absence of
the tip.  For example, the charged tip only pushes out the position of 
the secondary minimum in the density from $x/L=0.25$ to $x/L=0.3$.

In general, when we increase the tip potential to intermediate
values, the resolution of our micrographs improves compared with the
weak tip limit. This is evident in the fact that at $q=1.8e$ we can
already see signs of Wigner crystallization at $L=180nm$, in contrast
with the weak tip case with $q=0.02e$ where we can only detect
quasi-Wigner crystal at $L=250nm$. On the other hand, at $q=1.8e$ there
is \emph{no} signature of Wigner crystallization when the electrons are
in a liquid state in a $L=110nm$ wire. Thus we do not have "false
positive" signature of quasi-Wigner crystal, in contrast with the
strong tip limit described in Section\ref{strong}, where the 
micrographs show $N$ peaks for a $N$-electron wire regardless whether, 
in the wire in the absence of the tip, the electrons are in a liquid or 
a quasi-Wigner crystal state.

Because of these two characteristics, an intermediate tip
potential may help an experimentalist to reliably detect the presence
of a quasi-Wigner crystal state in a shorter, higher density wire
closer to the crossover from a liquid state.

\section{Single Electron}
Beyond the system of interacting electrons discussed above, a possible 
further application of Coulomb blockade microscopy is to experimentally 
``map'' the rugged potential landscape produced by wire inhomogeneities 
and charged impurities in the substrate. One would focus on the 
transition from $N=0$ to $N=1$ state, in which case the Coulomb 
blockade micrograph would reveal information about the single particle 
density.  By inverting the transformation in 
Eq.~\ref{equ:perturbation}, one may be able to approximately obtain the 
single particle ground state density $\rho(x)$.  In the absence of an 
external magnetic field, the ground state wavefunction $\psi(x)$ has no 
nodes and can be chosen to be $\psi(x)=(\rho(x))^{1/2}$  It is then
straightforward to invert the Schr\"odinger's equation to extract the 
potential landscape from the single particle wavefunction.

\section{Summary}
In summary, in this paper we show that tracking the peak position shift 
as a charged mobile tip moves above and across a nanowire, a technique 
we term Coulomb blockade microscopy, can reveal spatially-resolved 
information about the electronic density and states of a quantum 1D 
dot.  A weak tip potential can serve as a probe with a tunable width, 
to reveal the spatial distribution of the electronic density in the 
wire.  A strong tip potential that depletes part of the wire can be 
used to manipulate individual electrons from one partition to the 
other, and the accompanying Coulomb blockade micrograph can indicate 
the transitions between different partitionings.  Furthermore, a 
feature of the resulting micrograph, the distance between peaks marking 
the $(odd,odd)$ partitioning, can serve as an indicator of the relative 
strength of the interaction.

In this paper we have chosen extreme values of the tip charge $q$ to 
illustrate the physics in the two limits.  However, our calculations 
show that the discussions above hold true for a wider range of 
moderately small and large values of $q$.

To obtain a quantitative description of the energy shifts expected in
Coulomb blockade microscopy, particularly in the intermediate coupling
regime, we see that it is necessary to perform a realistic calculation,
which takes into account both the electron-electron interaction and the
non-linear effects of the charged tip on the electronic state of the wire.
If one is prepared to carry out such a calculation, however, Coulomb
blockade microscopy can be a powerful probe of interaction effects in
the wire.

\section{Acknowledgements}
We would like to thank Erin Boyd, Halvar Trodahl and Jesse Berezovsky 
and especially Bob Westervelt for helpful discussions. This work is 
supported in part by NSF grants PHY-0646094 and DMR-0906475. Numerical 
work was performed in part at the Center for Nanoscale Systems, a 
member of the National Nanotechnology Infrastructure Network (NNIN) supported 
by NSF award ECS-0335765. JQ is also supported in part by NIM and DFG through 
SFB 631.

\end{document}